\documentclass[aps,pre,twocolumn]{revtex4}
\usepackage{amsmath,bm,epsfig}

\def\Fbox#1{\vskip1ex\hbox to 8.5cm{\hfil\fboxsep0.3cm\fbox{%
  \parbox{8.0cm}{#1}}\hfil}\vskip1ex\noindent}  

\newcommand{\B}[1]{{\bm{#1}}}
\newcommand{\C}[1]{{\mathcal{#1}}}    
\let \= \equiv \let\*\cdot \let\~\widetilde \let\^\widehat \let\-\overline

\def\<{\left\langle}    \def\>{\right\rangle}
\def\({\left(}          \def\){\right)}
 \def \[ {\left [} \def \] {\right ]}

\begin{document}
\title{Predictive Statistical Mechanics for Glass Forming Systems}
\author{Laurent Bou\'e, Edan Lerner, Itamar Procaccia and Jacques Zylberg}
\affiliation{$^1$Department of Chemical Physics, The Weizmann Institute of Science, Rehovot 76100, Israel}
\
\date{\today}

\begin{abstract}
Using two extremely different models of glass formers in two and
three dimensions we demonstrate how to encode the subtle changes in
the geometric rearrangement of particles during the scenario of the
glass transition. We construct a statistical mechanical description
that is able to explain and predict the geometric rearrangement, the
temperature dependent thermodynamic functions and the
$\alpha$-relaxation time within the measured temperature range and
beyond. The theory is based on an up-scaling to proper variables
(quasi-species) which is validated using a simple criterion. Once
constructed, the theory provides an accurate predictive tool for
quantities like the specific heat or the entropy at temperatures
that cannot be reached by measurements. In addition, the theory
identifies a rapidly increasing typical length scale $\xi$ as the
temperature decreases.  This growing spatial length scale determines
the $\alpha$-relaxation time as $\tau_\alpha \sim
\exp(\mu\xi/T)$ where $\mu$  is a typical chemical potential per unit length.
\end{abstract}
\maketitle

\section{Introduction}

For typical glass formers with soft potentials the ``glass
transition'' is accompanied by a dramatic increase of relaxation
times: the viscosity of the super-cooled liquid shoots up by fifteen
orders of magnitude within a relatively short temperature range
\cite{96EAN}. When the liquid is examined on microscopic time
scales, the molecules seem jammed, trapped into moving only within
their cages. Of course, on the much longer viscous time scale these
same systems do reach equilibrium by restructuring themselves
through an ergodic exchange of particles between cages. As long as
the system is ergodic it lends itself, at least in principle
\cite{08EP}, to a statistical mechanical treatment. Nevertheless, it
is extremely hard to reach a useful statistical mechanical theory as
long as one considers the glass former on the level of its
constituent particles because it is difficult to evaluate
partition-function integrals in continuous coordinates. It is
therefore tempting to find a reasonable up-scaling (coarse-graining)
method that would define a discrete statistical-mechanics with
partition sums rather than integrals, with the sum running on a
finite number of quasi-species having well characterized
degeneracies and enthalpies. Indeed, in a number of examples in
2-dimensions (2D) \cite{07ABIMPS,07HIMPS,07ILLP,08LP,09LPR} and in
one example in 3D \cite{09LPZ} it was shown recently that such a
discrete statistical-mechanics is possible and quite advantageous
\cite{08HIP,08HIPS} in providing a successful description of the
statistics and the dynamics of systems undergoing the glass
transition. In this paper we examine how far the predictions of such
a statistical mechanical theory can be pushed. As the theory is
developed on the level of up-scaled quasi-species, one can ask
whether the energy, entropy and other thermodynamic quantities
appearing in the quasi-species language are identical numerically to
the corresponding quantities of the system itself when computed on
the level of constituting particles. One of the aims of this paper
is to answer this question in the affirmative. This means that the
statistical mechanical theory can be used to quantitatively predict
physical observables of interest also beyond the range of measurements,
hopefully providing a satisfactory understanding of the
phenomenology of glass formation.

To demonstrate the generality of the approach we examine two
extremely different models of glass formers to which we apply the
same procedure of: i)~identifying an appropriate up-scaling
ii)~validating that the chosen up-scaling yields a consistent
statistical mechanics and iii)~demonstrating how the resulting theory
provides a good understanding of the subtle geometric organizational
changes as a function of temperature, of the thermodynamic functions
and of the relaxation time. We will stress the point that
understanding the way that the concentrations of the available
up-scaled quasi-species depend on the temperature is tantamount to
understanding the scenario of the glass transition. To underline the
generality of the approach we deal below with two models, one model
is studied in 3D in an NPT ensemble and the other in 2D in an NVT
ensemble. A short announcement of the results pertaining to the
first model was published in \cite{09LPZ} and here we provide
considerably richer details. The results of the second model are
novel to the present paper. The models were chosen, as explained
below, to represent extremely different microscopic characteristics;
nevertheless once up-scaled, the statistical mechanics appears very
similar encouraging us to propose that the approach is rather general.

The structure of the paper is as follows: in Sect. \ref{models} we
introduce the two models studied below, discuss the dynamics of
their correlation functions as a function of temperature, and
measure the relaxation time $\tau_\alpha$ that will be later
connected to the statistical mechanical theory. In Sect.
\ref{upscaling} we present the upscaling for these two models. This
upscaling defines quasi-species in terms of carefully chosen groups
of particles rather than individual particles. We then show how the
scenario of the glass formation can be characterized by the
temperature dependence of the concentrations of the various
available quasi-species. Since this upscaling is not unique, we
present a {\em validation} of the choice of quasi-species. The
validation is a demonstration that the choice of quasi-species
provides a self-consistent statistical mechanics. In other words,
each quasi-species is characterized by its own enthalpy and its own
degeneracy. The statistical mechanics is then shown to be able to
accurately recapture the scenario of the glass formation in
temperatures where data is available, and to continue to predict it
for temperatures that are outside the range of numerical
simulations. In Sect. \ref{dynamics} we relate the upscaled picture
to the dynamics of the system. We show how a natural static
length-scale is appearing, and how this length scale characterizes
the relaxation time $\tau_\alpha$. Finally, in Sec. \ref{thermo} we
demonstrate that the statistical mechanics as defined on the
upscaled quasi-species is able to provide the correct thermodynamic
function for the original system. In other words, we can compute the
energy, entropy, specific heat etc. directly from the upscaled
picture, in the range of temperatures accessible to simulations as
well as for temperatures that are inaccessible to simulations. In
Sect. \ref{summary} we discuss the degree of generality of the
present approach and stress what are the remaining riddles for
future research.

\section{Two models of Glass Formation}
\label{models}

\subsection{Three-dimensional Binary Model in NPT ensemble}

\begin{figure}
\hskip -1.2cm
\centering
\includegraphics[scale = 0.50]{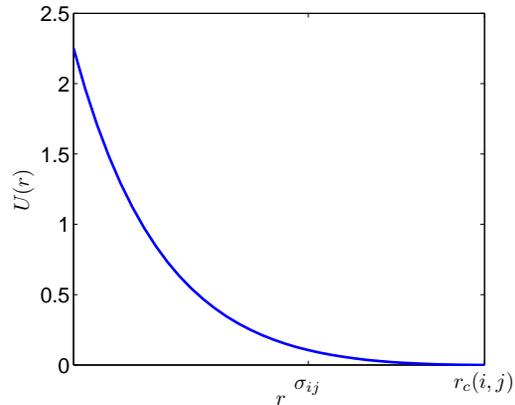}
\caption{Color online: The pair-wise potential of the binary model}
\label{potbinary}
\end{figure}

Some results concerning the upscaling of the first model that is
studied below were announced already in \cite{09LPZ}, and here we
provide further details. The model is a version of a much studied
model \cite{93KA,89DAY,99PH,08BPGSD}, here of a 50:50 mixture of $N$
point-particles in 3-dimensions ($N=4096$ in our case), interacting
via a binary potential. We refer to half the particles as `small'
and half as `large'; they interact via the potential $U(r_{ij})$:
\begin{equation}
\label{potential}
U(r_{ij}) =
\left\{
\begin{array}{ccl}
\epsilon\left[ \left(\frac{\sigma_{ij}}{r_{ij}} \right)^\alpha
- \left(\frac{\sigma_{ij}}{r_{ij}} \right)^\beta + a_0 \right] & , & r_{ij} \le r_c(i,j) \\
0 & , & r_{ij} > r_c(i,j)
\end{array}
\right.
\end{equation}
Here, $\epsilon$ is the energy scale and $\sigma_{ij} = 1.0\sigma, 1.2\sigma$ or
1.4$\sigma$ for small-small, small-large or large-large interactions,
respectively. For the sake of numerical speed the potential is
cut-off smoothly at a distance, denoted as $r_c$, which is
calculated by solving $\partial U/\partial r_{ij}|_{r_{ij} = r_c} =
0$ which translates to $r_c =
\left({\alpha}/{\beta}\right)^{\frac{1}{\alpha-\beta}}\sigma_{ij}$.
The parameter $a_0$ is chosen to guarantee the condition $U(r_c) =
0$. Below we use $\alpha=8$ and $\beta=6$, resulting in $r_c =
\sqrt{8/6}~\sigma_{ij}$ and $a_0 = 0.10546875$, see Fig.
\ref{potbinary}. Note that the potential is purely repulsive, in a
premeditated distinction from the second model discussed in Subsect.
\ref{hump}

\begin{figure}
\hskip -1.2cm
\centering
\includegraphics[scale = 0.45]{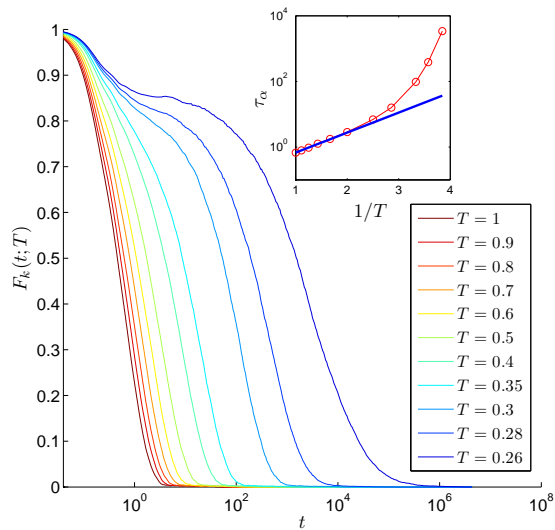}
\caption{Color online: Time dependence of the correlation functions
(\ref{defFk}) for a range of temperatures (decreasing from left to
right) as shown in the figure. The inset shows the relaxation time
$\tau_\alpha$ in a log-lin plot vs $1/T$, compared to an Arrhenius
temperature dependence.} \label{corr}
\end{figure}

The model dynamics were studied as a function of the temperature
keeping the pressure fixed at $p=10$ (NPT ensemble). The model was simulated by employing the Verlet integration scheme with the Berendsen thermostat and barostat \cite{91AT}. The units of mass are the mass of the particles $m$,
energy is measured in units of $\epsilon$ and the fundamental length scale is $\sigma$.
The slowing down in the super-cooled regime is exemplified by measuring the
self-part of the intermediate scattering function \cite{99PH} summed
over the large particles only,
\begin{equation}
F_k(t;T) \equiv \left\langle \case{2}{N}\sum_{i=1}^{N/2} \exp\left\{{i\B k \cdot [\B r_i(t)-\B r_i(0)]}\right\}\right \rangle  \ . \label{defFk}
\end{equation}
In Fig. \ref{corr} we show these correlation functions for
$k=5.1\sigma^{-1}$ and for a range of temperatures as indicated in
the figure. We see the usual rapid slowing down that can be measured
by introducing the typical time scale $\tau_\alpha$ that is
determined by the time where $F_k(t=\tau_\alpha;T)= F_k(0;T)/e
\equiv 1/e$. The relaxation times are shown in the inset of Fig.
\ref{corr} as a function of $1/T$ in a log-lin plot to stress the
non-Arrhenius dependence at lower $T$.

\subsection{Two-dimensional `hump' model in NVT ensemble}
\label{hump}

The second model is constructed following \cite{89Dzu} such as to
have very different microscopic properties from the binary model.
The interaction potential is constructed as a piecewise function
consisting of the repulsive part of a standard 12-6 Lennard--Jones
potential connected at $r_0 = 2^{1/6}\sigma$ to a polynomial
interaction $P(x) = \sum_i a_i x^i$.  The $a_i$'s are tuned so that
$P(x)$ displays a peak at $r = r_{\mbox{\tiny hump}}$ and also such
that there is a smooth continuity (up to second derivatives) with
the Lennard--Jones interaction at $U(r_0) = \epsilon h_0$ as well
as with the cut-off interaction range $U(r_{\star}) = 0$.  The
interaction potential for the hump model reads:
\begin{equation}
U(r_{ij})=\left\{
\begin{array}{c @{,}c}
\epsilon \left[ \left( \frac{\sigma}{r_{ij}}  \right)^{12} - \left( \frac{\sigma}{r_{ij}} \right)^6 + \frac{1}{4} + h_0 \right] & $\quad$ r_{ij} \leq r_0 \\
\epsilon h_0 P \left( \frac{r_{ij} - r_0}{r_{\star} - r_0} \right)  & $\quad$  r_0 < r_{ij} \leq r_{\star} \\
0 &  $\quad$ r_{ij} > r_{\star}
\end{array}
\right.
\label{humpot}
\end{equation}
\begin{figure}
\begin{center}
\includegraphics[width=0.8\linewidth]{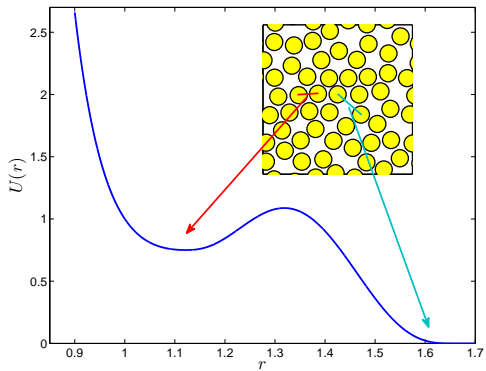}
\caption{The pair-wise potential for the hump model. In the inset we show a snapshot of the position
of the point-particles (the circles represent the radius $\sigma$). Note that the two typical scales $r_{\rm min}$ and
$r_\star$ appear as the typical distances between particles}.
\label{fighump}
\end{center}
\end{figure}
with the parameters as shown in table 1.
\begin{table}
\begin{center}
\begin{tabular}{|l|r@{.}l|}
\hline
$h_0  $ & 0 & 75 \\
$r_{\mbox{\tiny hump}}  $ & 1 & 32 \\
$r_{\star}$ & 1 & 65 \\
\hline
$a_0  $ & 1 & 0 \\
$a_1  $ & 0 & 0 \\
$a_2  $ & 2 & 675405732987203 \\
$a_3  $ & 46 & 593574934286437\\
$a_4  $ & -212 & 021700143354 \\
$a_5  $ & 308 & 7495934944721 \\
$a_6  $ & -188 & 1854149609638 \\
$a_7  $ & 41 & 188540942572089 \\
\hline
\end{tabular}
\caption{{\it Parameters used in the potential of the hump model.}}
\end{center}
\end{table}

Note that the two typical distances that are defined by this potential, i.e. the distance at the minimum $r_{\rm min}$ and the cutoff scale $r_\star$, appear
explicitly in the amorphous arrangement of the particles in the supercooled liquid, as shown in the inset in Fig.~\ref{fighump}. The model has two crystalline ground states, one at high pressure with a hexagonal lattice and a lattice constant of the order
of $r_{\rm min}$. At low pressure the ground state is a more open structure in which the distance $r_{\star}$ appears periodically. At intermediate pressures the system fails to crystallize and forms a glass upon cooling.

\begin{figure}
\hskip -1.0cm \centering
\includegraphics[scale = 0.40]{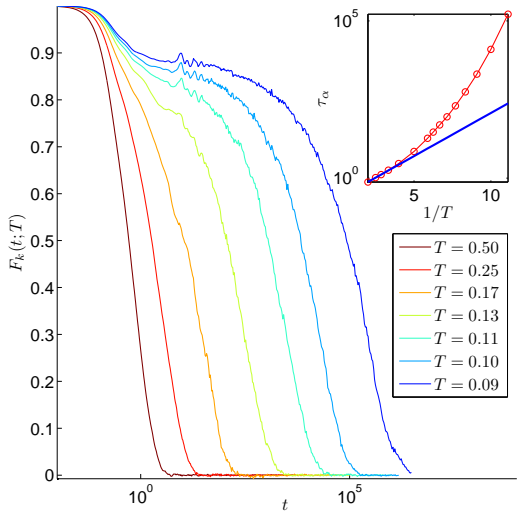}
\caption{Color online: Time dependence of the correlation function
\ref{defFhump} in the hump model for a range of
temperatures as shown in the figure. Inset: the temperature
dependence of the relaxation time $\tau_\alpha(T)$.}. \label{cortauhump}
\end{figure}

The great difference between these two models will become even
clearer after we define the up-scaled quasi-species below. On the
face of it, the scenario for the glass transition appears similar as
exemplified in Fig. \ref{cortauhump} in which the intermediate
scattering function
\begin{equation}
F_k(t;T) \equiv \left\langle \case{1}{N}\sum_{i=1}^{N} \exp\left\{{i\B k \cdot [\B r_i(t)-\B r_i(0)]}\right\}\right \rangle  \ . \label{defFhump}
\end{equation}
is shown for $k=6.16\sigma^{-1}$. The hump model was simulated using the same algorithms as the binary model, but without
the Berendsen barostat that was unneeded for the NVT ensemble.

The relaxation time $\tau_\alpha$ for the hump model is shown in
Fig. \ref{cortauhump} in which we stress again the difference
between the region in which there is Arrhenius behavior and the
region where the relaxation time grows faster than $\exp (E/T)$.

\section{Choice of up-scaled variables and validation}
\label{upscaling}

Our aim is to provide a theory that captures, when the temperature goes down, the subtle changes in the structural organization of the particles; this rearrangement is in fact directly responsible for the glass transition. The first step in our approach is up-scaling, (or coarse-graining) from particles to quasi-species that can be characterized by their enthalpy and degeneracy.  Up-scaling is an art, since it can be done in various ways and there is no unique algorithm to select a-priori a `best' up-scaling. Therefore we need to validate the choice of up-scaled variables using a criterion that was introduced in \cite{09LPZ}. The up-scaling is typically different for different models, and we describe it separately for the two models at hand.
\subsection{Up-scaling of the binary model}

\begin{figure}
\hskip -1.0cm
\includegraphics[scale = 0.435]{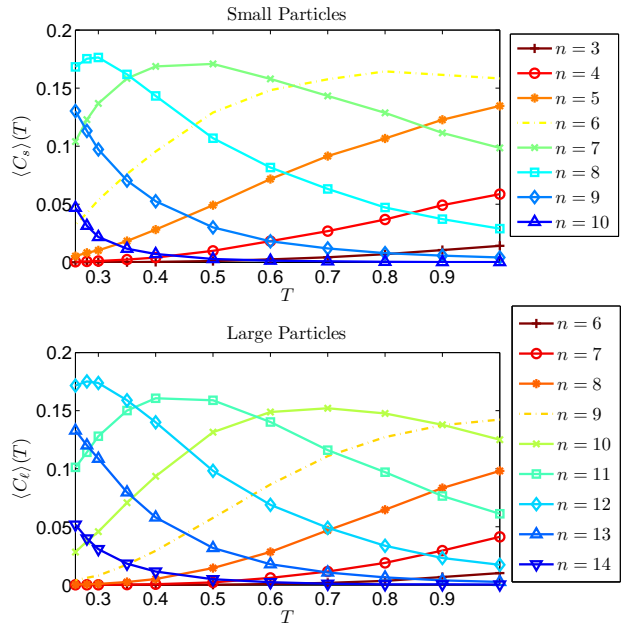}
\caption{Color online: Temperature dependence of the concentrations
of the quasi-species of the binary model. Symbols are simulation
data and the lines are a guide to the eye. } \label{conc}
\end{figure}

The potential in the case of the binary model is purely repulsive
and it has well defined range of interaction $r_c(i,j)$ for any
given pair of particles. A natural up-scaling for this potential is
provided by the particles and their nearest neighbors, where
`neighbors' are defined as all the particles $j$ around a chosen
central particle $i$ that are within the range of interaction
$r_c(i,j)$. We refer to the type of a central particle (small or large)
in combination with the amount of its nearest neighbors to define
the quasi-species. In the interesting range of temperatures we find
8 quasi-species with one `small' central particle and $3,4\dots 10$
neighbors, and 9 quasi-species with one `large' central particle
with $6,7\dots 14$ neighbors, all in all 17 quasi-species. Other
combinations have negligible concentration ($<0.5\%$) throughout the
temperature range. We denote these quasi-species as $C_s(n)$ and
$C_\ell(n)$ with $s$ and $\ell$ denoting the small or large central
particle, while $n$ denotes the number of neighbors. We measured the
mole-fractions $\langle C_s(n)\rangle(T)$ and $\langle
C_\ell(n)\rangle(T)$ and the results are shown in Fig. \ref{conc}.
We see that some concentrations decrease upon decreasing the temperature, others increase, and yet some first increase and further decrease. We submit to the reader that the subtle changes in configurational arrangement as seen in this plot
encode the scenario of the glass transition in a way that we will attempt to decode below.
\subsection{Up-scaling of the hump model}

\begin{figure}
\hskip -0.0cm
\includegraphics[scale = 0.42]{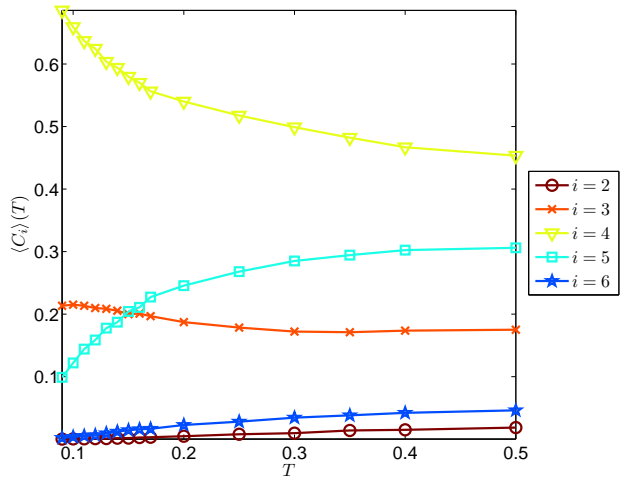}
\caption{Color online: Temperature dependence of the concentrations
of the quasi-species of the hump model. Symbols are simulation data
and the lines are a guide to the eye. } \label{humpconc}
\end{figure}

The potential of the hump model has a distinct minimum which
suggests that quasi-species could consist of each particle
and all its neighbors which are within the minimum. This type of
up-scaling may have a limited usefulness when the temperature is
sufficiently high to allow easy escape over the maximum. The glass
transition occurs however in a temperature range where $T$ is lower
than the maximum, and we expect the chosen up-scaling to be sensible
throughout the interesting temperature range. In this range we find
quasi-species with 2, 3, 4, 5 and 6 neighbors within the minimum. In
Fig. \ref{humpconc} we show the temperature dependence of the
quasi-species of this model which are denoted as $C_i$ with
$i=2,3,\dots,6$ being the number of neighbors within the minimum.
The reader should be sensitive to the distinction between the two
models. The quasi-species in the binary model can change upon every
cage vibration; anytime the distance between two particles exceeds
or goes below $r_c$ the definition of the quasi-speices changes. On
the other hand in the hump model the quasi-species are much more
stable, since an exchange of a particle calls for an escape or
penetration by going over the hump. We have chosen the models to
have these clear differences in order to test the generality of our
approach.

\subsection{Validation of the up-scaling}

Since there is no unique algorithm to choose the up-scaling, we need
to have a criterion that validates or rejects the choice of
quasi-species. In ref. \cite{09LPZ} such a criterion was introduced
as explained next.

\subsubsection{Validation of the up-scaling in the binary model}

\begin{figure}
\hskip -1.0cm
\centering
\includegraphics[scale = 0.40]{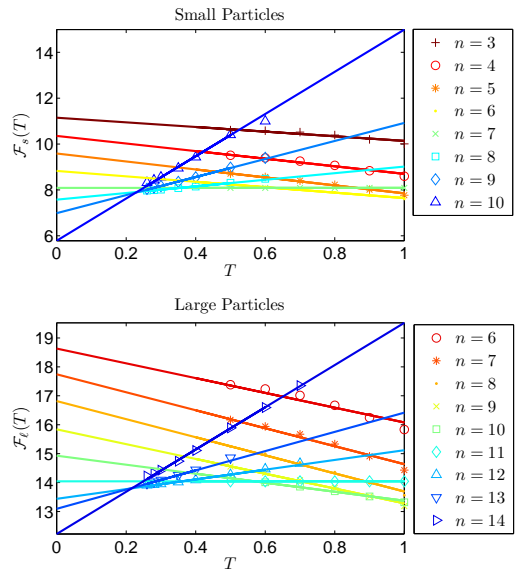}
\caption{Color online: The approximate linear dependence of the free energies of the chosen quasi-species on the temperature.
From the slope we read the degeneracy and from the intercept the enthalpies (up to normalization), cf. Eq. \ref{readHg}. Note that when the free energies are large we do not have data: the concentrations become exponentially small and in a finite simulation box they disappear completely.}
\label{linear}
\end{figure}

\begin{figure}
\hskip -0.5 cm
\includegraphics[scale = 0.45]{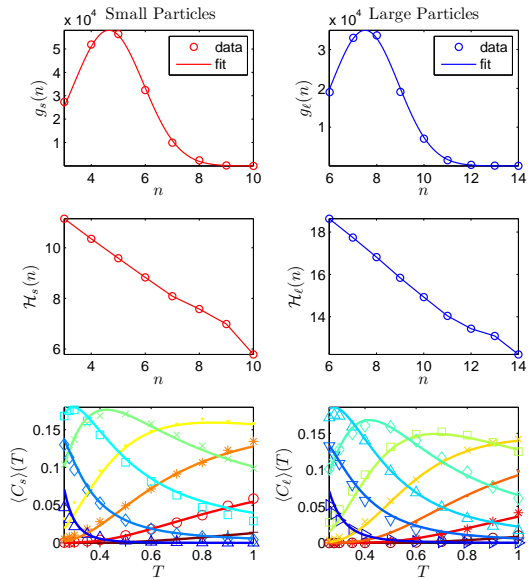}
\caption{Color online: Upper panel: The degeneracies $g_s(n)$ and $g_\ell(n)$ read from the slopes of Fig. \ref{linear} (in circles) and the degeneracies according to the gaussian model Eq. (\ref{comparisons}). Middle panel: the measured enthalpies. Lower panel: comparison of the measured concentrations of quasi-species to those calculated from Eqs. (\ref{DefgH}) using the model degeneracies and measured enthalpies. Here symbols are data and lines are theoretical predictions.}
\label{comparisons}
\end{figure}

To decide whether the up-scaling provides a useful statistical
mechanics for the binary model we now ask whether there exist free
energies $\C F_s(n;T)$ and $\C F_\ell(n;T)$ such that
\begin{eqnarray}
\langle C_s(n)\rangle (T)& =& \frac{ e^{-\C F_s(n;T)/T}}{2\sum_{n=3}^{10}  e^{-\C F_s(n;T)/T}} \ ,\nonumber\\
\langle C_\ell(n)\rangle (T)& =& \frac{e^{-\C F_\ell(n;T)/T}}{2\sum_{n=6}^{14} e^{-\C F_\ell(n;T)/T}} \ . \label{DefF}
\end{eqnarray}
The free energies are found by inverting Eqs. (\ref{DefF}) in terms
of the measured concentrations. In doing so one can always choose
{\bf one} of the concentration to have by definition zero free
energy. This is because the constraint $\sum_n C_s(n)+\sum_n
C_\ell(n)=1$ makes the system of equations overdetermined. Then the
free energies of all the other quasi-species are actually the
difference from this chosen reference particle. We then plot these
quantities as a function of the temperature, as demonstrated for the
present case in Fig. \ref{linear}.
We say that our upscaling is validated if $\C F_s(n;T)$ and $\C
F_\ell(n;T)$ can be well approximated as linear in the temperature.
Then we can interpret
\begin{eqnarray}
\C F_s(n;T) &\equiv& \mathcal H_s(n) - T \ln g_s(n)\nonumber \ , \nonumber\\
\C F_\ell(n;T) &\equiv& \mathcal H_\ell(n) - T \ln g_\ell(n) \ ,
\label{readHg}
\end{eqnarray}
where now the degeneracies $g_s(n)$ and $g_\ell(n)$ (read from the
slopes in Fig. \ref{linear}) and enthalpies $\C H_s(n)$ and $\C
H_\ell(n)$ (read from the intercepts) are {\bf
temperature-independent}. This validates the choice of up-scaling.
In other words, the approximate linearity of the inverted free
energies in the temperature means that we can write the
concentrations as
\begin{eqnarray}
\langle C_s(n)\rangle (T)& \approx& \frac{g_s(n) e^{-\mathcal H_s(n)/T}}{2\sum_{n=3}^{10} g_s(n) e^{-\mathcal H_s(n)/T}} \ ,\nonumber\\
\langle C_\ell(n)\rangle (T)&\approx& \frac{g_\ell(n) e^{-\mathcal
H_\ell(n)/T}}{2\sum_{n=6}^{14} g_\ell(n) e^{-\mathcal H_\ell(n)/T}}
\ . \label{DefgH}
\end{eqnarray}
Then we can use these forms also as a prediction for temperatures
where the simulation time is too short to observe the relaxation to
equilibrium. The resulting degeneracies $g_s(n)$ and $g_\ell(n)$ can
be easily modeled by a Gaussian
distribution around the most probable number $n_{\rm mp}$ of nearest
neighbors for small and large particles respectively:
\begin{eqnarray}
g_s(n)&\approx& e^{-[(n-n^s_{\rm mp})^2/2\sigma^2_s ]} \ , \quad n^s_{\rm mp}=4.65, \sigma^2_s=1.55\ , \nonumber\\
g_\ell(n)&\approx& e^{-[(n-n^\ell_{\rm mp})^2/2\sigma^2_\ell ]}\ , \quad n^\ell_{\rm mp}=7.50,  \sigma^2_\ell=2.0 \ . \label{gasussian}
\end{eqnarray}
The analytic fit to the measured degeneracies is
shown in Fig. \ref{comparisons}, upper panel. The same figure shows
in the middle panel the enthalpies of the various quasi-species. One
could model the enthalpies as a linear function in $n$. These
results are easily interpreted; we have high enthalpies when there
are large free volumes (few neighbors). The lowest enthalpies are
found when there are many neighbors and there is no much costly free
volume. In other words, at the present density and range of
temperatures the $pV$ term dominates the energy in the enthalpy.
Using the theoretical degeneracies and the measured enthalpies we
compute the concentrations of all our quasi-species and compare them
with the measurement in the lowest panel of Fig.~\ref{comparisons}.
The agreement that we have, especially considering the number of
quasi-species and the simplicity of the theory, is very
satisfactory. Notice that the competition between degeneracy and
enthalpy explains the rather intricate temperature-dependence of the
concentrations of the quasi-species, sometimes declining when the
temperature drops, sometime rising, and sometime having
non-monotonic behavior.
\subsubsection{Validation of the up-scaling in the hump model}

\begin{figure}
\hskip -.5cm \centering
\includegraphics[scale = 0.45]{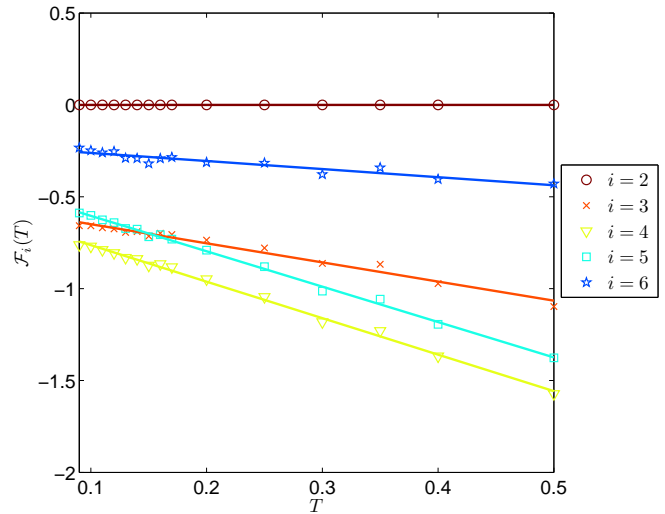}
\caption{Color online: The approximate linear dependence of the free energies of the chosen quasi-species on the temperature
in the hump model.
From the slope we read the degeneracy and from the intercept the energies(up to normalization), cf. Eq. \ref{humpF}.}
\label{humplin}
\end{figure}

Validating the up-scaling in the hump model follows the same ideas
as in the binary model. We choose the quasi-species $n=2$ as the
reference concentration with $F_2\equiv 0$ and then invert the data
for $\langle C_i\rangle(T)$ to find the free energies via
\begin{equation}
\langle C_i\rangle(T) = \frac{e^{-\mathcal F_i(T)/T}}{\sum_2^6
e^{-\mathcal F_i(T)/T} } \ .
\end{equation}
The observed linearity of $\mathcal F_i(T)$ in the temperature means
that we can write
\begin{equation}
\mathcal F_i(T) = \mathcal H_i -T \ln g_i \ , \label{humpF}
\end{equation}
where $\C H_i$ is the enthalpy and $g_i$ the degeneracy of the $i$th
quasi-specie. The test of linearity from which we can read the
energies and degeneracies is shown in Fig. \ref{humplin}.
From the intercepts of the lines in Fig.~\ref{humplin} we read the enthalpies, and from the slopes we read the degeneracies.
The resulting numerical values are shown in Table \ref{tablehump}.
\begin{table}
\begin{center}
\begin{tabular}{|l|l||l|l|}
\hline
$\mathcal{H}_2$ & 0. & $g_2$ & 1. \\
$\mathcal{H}_3$ & -0.545053 & $g_3$ & 2.830972 \\
$\mathcal{H}_4$ & -0.564547 & $g_4$ & 7.289318 \\
$\mathcal{H}_5$ & -0.411079 & $g_5$ & 6.862693 \\
$\mathcal{H}_6$ & -0.217412 & $g_6$ & 1.552314 \\
\hline
\end{tabular}
\caption{ Temperature independent enthalpies and degeneracies of
the sub-species of the hump model. } \label{tablehump}
\end{center}
\end{table}

Since the hump model is studied in the NVT ensemble the reader may
ask why we get enthalpies rather than energies. The reason is that the
upscaling using the index $i$ takes into account only the particles that
reside inside the hump. The energy of a quasi-species with $i$ particles within
the hump is proportional to $i$, and thus for low temperatures the energetically preferred
subspecies would be the one with no neighbors within the hump. Clearly this arrangement
does not satisfy the constant volume constraint which needs to be taken into account.
Indeed, what we call enthalpies
cannot be straight energies, otherwise the energies should have been
proportional to $i$, making $\C H_2$ the smallest rather than the
largest.

\subsection{Summary of the statistical mechanics}

We can summarize the findings up to now by saying that at least in
terms of capturing the scenario of the changes in the spatial
organization of the particles our statistical mechanics appears very
suitable. Our quasi-species appear to have very well defined
enthalpies and degeneracies and we can therefore capture the
temperature dependence of the concentrations of the quasi-particles
with high accuracy. Our thesis is that this scenario is actually the
scenario of the glass transition, and hidden in it is also the
reason for the slowing down, as demonstrated in the next section.
Intuitively the picture should be clear already now. As the
temperature reduces the quasi-species with high free energy tend to
disappear, leaving us with quasi-species of low free energy. Then
the dynamics become more constrained, since it is more costly to
change objects of low free energy (creating on the way objects of
higher free energy) than at high temperature when there are many
objects with high free energy that can readily change to objects of
lower free energy. What we need now is to make this observation more
quantitative.

\section{Relation to Dynamics}
\label{dynamics} In this section we connect the structural theory to
the dynamical slowing-down. To this aim we note that in both models
there are a number of quasi-species whose concentration goes down
exponentially (or maybe faster) when the temperature decreases, and
that the relaxation time shoots up at the same temperature range. We
refer to these quasi-species as the `liquid' ones; These are the
quasi-species that are prevalent when the temperature is high but
are getting rare when the temperature goes down. We discuss now the
connection between this phenomenon and the slowing down.

\subsection{The binary model}

\begin{figure}
\hskip -1. cm
\centering
\includegraphics[scale = 0.4]{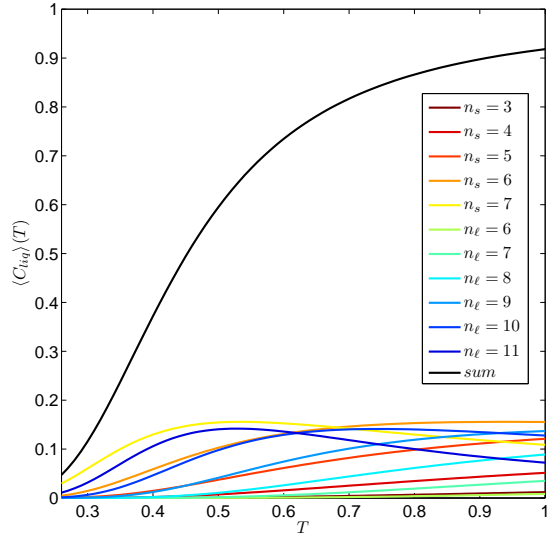}
\caption{Color online: The temperature dependence of $C_{\rm liq}(T)$ is shown as the upper continuous line. The contributions of the
various `liquid' sub-species are shown with symbols which are identified in the inset.}
\label{cliq}
\end{figure}

In this example the `liquid' concentrations are those consisting of small particles with 3-7 neighbors, and large particles with 6-11 neighbors, see Fig. \ref{cliq} \cite{footnote}. We sum up these concentrations and denote the sum as $\langle C_{\rm liq}\rangle (T)$. The dependence of $\langle C_{\rm liq}\rangle (T)$ on the temperature is shown in Fig. \ref{cliq}.
This concentration is used to define
a typical scale,
\begin{equation}
\xi(T) \equiv [\rho C_{\rm liq}(T)]^{-1/3} \ ;
\end{equation}
where $\rho$ is the number density. This length scale has the physical interpretation of the average distance between the `liquid' quasi-species. It was argued before \cite{07ILLP,08LP,08EP} that this length scale can be also interpreted as the linear size of relaxation events which include $O(\xi(T))$ quasi-species. We can therefore estimate the growing free energy per relaxation event as $\Delta G=\mu \xi(T)$ where $\mu$ is the typical chemical
potential per involved quasi-species. This estimate, in turn, determines the relaxation time as
\begin{equation}
\tau_\alpha(T)= e^{\mu\xi(T)/T} \ . \label{fit}
\end{equation}
\begin{figure}
\hskip -1.0cm \centering
\includegraphics[width=200pt]{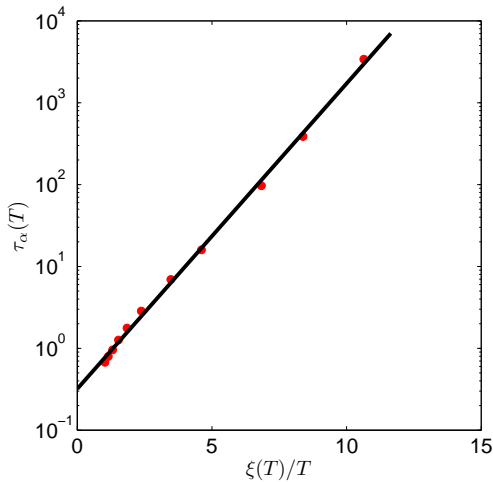}
\caption{Color online: The relaxation time $\tau_\alpha(T)$ in terms
of the typical scale $\xi(T)$ in the binary model. We show the
excellent fit to Eq. (\ref{fit}) with $\mu=0.37$. Note that the
intercept at $T\to \infty$ is of the order of unity as it must be.}
\label{tauvsxi}
\end{figure}
The quality of this prediction can be gleaned from
Fig.~\ref{tauvsxi}, where we can see that the fit is excellent, with
$\mu\approx 0.3$. The intercept in Fig. \ref{tauvsxi} is of the
order of unity; this is very reassuring, since this is what we
expect when $T\to \infty$.

\subsection{The hump model}

\begin{figure}
\hskip -1.2cm
\centering
\includegraphics[width=200pt]{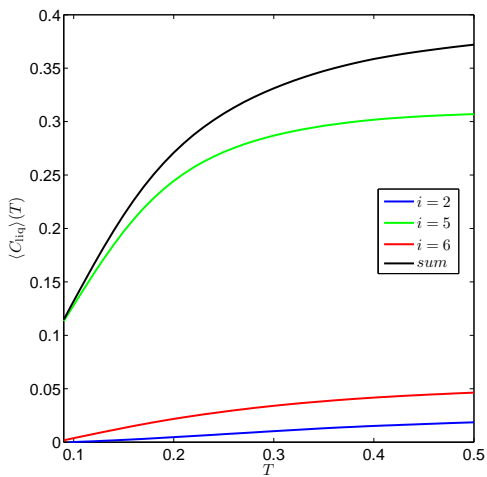}
\caption{Color online: The quasi-species contributing to $C_{\rm
liq}(T)$ in the hump model, and the temperature dependence of the
their sum, which is $C_{\rm liq}(T)$.} \label{humpliq}
\end{figure}

\begin{figure}
\hskip -1.2cm
\centering
\includegraphics[width=200pt]{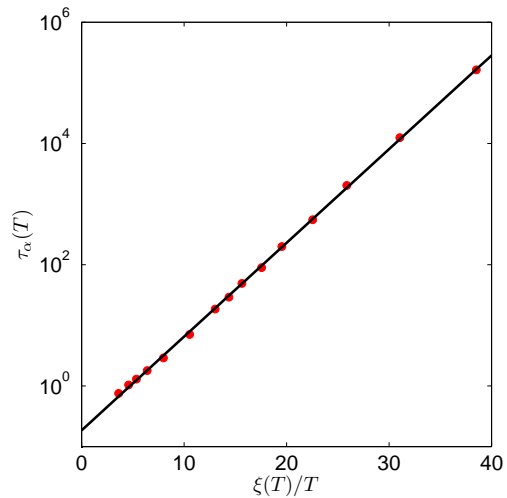}
\caption{Color online: The relaxation time $\tau_\alpha(T)$ in terms
of the typical scale $\xi(T)$ in the hump model. We show the
excellent fit to Eq. (\ref{fithump}) with $\mu=0.34$. Note that the
intercept at $T\to \infty$ is of the order of unity as it must be.}
\label{tauvsxi2}
\end{figure}

In this model the quasi-species whose concentration goes to zero rapidly in the relevant temperature range are those
with $i=2,5$ and 6. In Fig. \ref{humpliq} we see their temperature dependence and also their sum which is again denoted
as $C_{\rm liq}(T)$. Since in this model we are in two dimensions, the typical scale $\xi(T)$ is related to $C_{\rm liq}(T)$
according to
\begin{equation}
\label{fithump} \xi(T) \equiv [\rho C_{\rm liq}(T)]^{-1/2} \ ;
\end{equation}
With this obvious change we expect Eq. (\ref{fit}) to be valid also here, and indeed in Fig. \ref{tauvsxi2}
we see that this expectation is wonderfully fulfilled.

\subsection{Pertinent Remarks}
A few points should be stressed. As we expect (cf. Ref \cite{08EP}), in systems with point particles and soft potential, there is no reason to fit the relaxation time to a Vogel-Fulcher form \cite{96EAN} which predicts a singularity at finite temperature. In our approach
we predict that $\xi\to \infty$ only when $T\to 0$, and there is nothing singular on the way, only slower and slower relaxation. At some point the simulation time will be too short for the system to relax to equilibrium, but we can use Eq. (\ref{fit}) to predict what should be the simulation time to allow the system to reach equilibrium. It is important to bear in mind that
finite systems of point particles with soft potential are different from finite granular media or systems of hard spheres which can truly jam and lose ergodicity. Point particles with soft potential remain ergodic \cite{08EP}, and therefore should be amenable in their super-cooled regime to statistical mechanics. We argue that in order to construct simple, workable statistical mechanics one needs to up-scale the system and find a collection of quasi-species with well defined enthalpies and degeneracies. In this paper we showed that the simple criterion to validate the choice of up-scaling which was suggested
in \cite{09LPZ} can be applied to very different models once we select the proper up-scaling. Once the structural theory is under control, a natural length scale appears and can be used to determine the relaxation time, also for temperatures that cannot be simulated due to the fast growth of the relaxation time. The fact that the present approach works equally well in two and three dimensions, in NPT and NVT ensembles and in very different models  provides good reason to
believe that it has a substantial degree of generality.
\section{The thermodynamics of glass forming systems}
\label{thermo}
As a last issue we raise the following question: is the statistical mechanical theory as defined on the up-scaled
quasi-species determining also the thermodynamics of the original system on the particle level. The answer to this
question is not obvious a-priori, but turns out to be in the affirmative. As done throughout this paper, we demonstrate
this point separately for the two models at hand.

\subsection{Thermodynamics of the binary model}

\begin{figure}
\hskip -1.cm
\centering
\includegraphics[scale = 0.43]{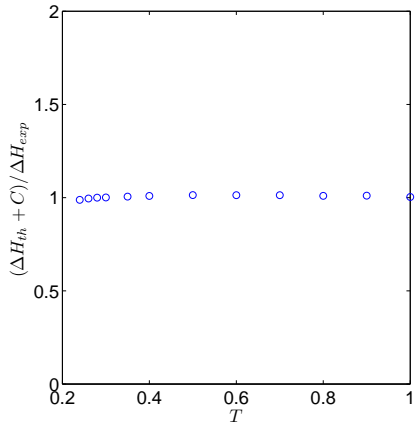}
\caption{Color online: The ratio of the LHS of Eq. (\ref{yes}),
denoted as $\Delta H_{\rm th} +C$ (here, $C=2.8$) over the RHS of
Eq. (\ref{yes}), denoted $\Delta H_{\rm exp}$. We conclude that the
enthalpy of the system is excellently reproduced by the up-scaled
variables.} \label{compbin}
\end{figure}

The binary model is studied in the NPT ensemble and therefore the first question to ask is whether the average enthalpy
of the quasi-species represents the correct enthalpy of the system. In other words, we have extracted the
enthalpies $H_s(n,T)$ and $H_\ell(n,T)$. Is it then true that the enthalpy of the system, computed as the average potential
energy summed over all the pairs of particles and summed with $pV$ should equal the enthalpy of the quasi-species:
\begin{eqnarray}
\sum_{n=3}^{10} C_s(n) \C H_s(n,T&)&+\sum_{n=6}^{14} C_\ell(n) \C H_\ell(n,T) \nonumber\\
&= &\frac{1}{2N}\sum_{i,j=1}^N \langle U(r_{ij})\rangle +pV(T) \ .
\label{yes}
\end{eqnarray}
In answering this question we remember that our enthalpies are defined up to an arbitrary constant since we chose
one of the quasi-species as a reference with zero free energy. We can therefore add or subtract an arbitrary constant
from the LHS or the RHS of Eq. \ref{yes}. In Fig. \ref{compbin} we show the ratio of the LHS of Eq. (\ref{yes}) to the
RHS, with the conclusion that the enthalpy of the system is excellently reproduced by the up-scaled variables.
Needless to say, we could therefore compute the specific heat at constant pressure, $C_p$, from either set of data,
and from that we could get the entropy of the system, reassuring us that the up-scaling method, once validated
as above, provides also the correct thermodynamics for the glass forming system. In the next subsection we show
actual computations of such thermodynamic quantities.
\subsection{Thermodynamics of the hump model}

\begin{figure}
\hskip -1.2cm
\centering
\includegraphics[scale = 0.43]{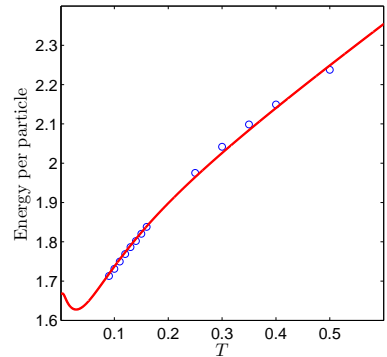}
\caption{Color online: Comparison of the LHS of Eq. (\ref{yes2}) in red continuous line,
to the RHS of the same equation, in symbols.  We conclude that the
energy of the system is excellently reproduced by the up-scaled
variables down to temperatures of about $T\approx 0.05$ For lower temperatures the extrapolation
obviously fails.} \label{comphump}
\end{figure}

\begin{figure}
\hskip -1.2cm
\centering
\includegraphics[scale = 0.43]{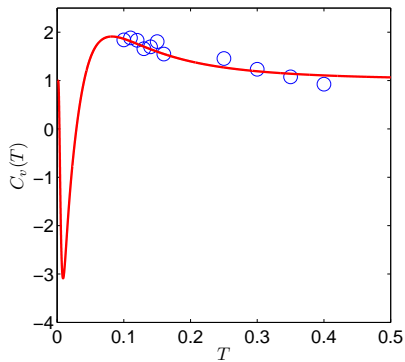}
\caption{Color online: Symbols: the derivative with respect to $T$
of the RHS in Eq. (\ref{yes2}). Continuous line: derivative with
respect to $T$ of the LHS in Eq. (\ref{yes2}). The theory predicts the expected specific heat peak and fails
when the specific heat becomes negative.} \label{compspec}
\end{figure}

The hump model was studied in the NVT ensemble, and therefore it is interesting to test whether one obtains the correct system's energy.  The potential energies of the quasi-species which were
characterized by the number of neighbors that are bound within the minimum are to an excellent approximation linear in that number, $E_i=0.83 i$ as measured with respect to the same zero point as that of the potential (\ref{humpot}).
We therefore need to check whether
\begin{equation}
\frac{0.83}{2}\sum_{i=2}^6 C_i(T) i + T  = \frac{1}{2N}
\sum_{i,j=1}^N  \langle U(r_{ij}) \rangle \ . \label{yes2}
\end{equation}
We have added $T$ for the excitation around the ground state of the quasi-species (two degrees of freedom in 2D). This test is shown in Fig. \ref{comphump} where the predictions of the theory were extrapolated all the way
to zero temperature. Obviously the data loses its predictability somewhere around $T=0.05$ where the specific heat
becomes negative. This artifact is seen even better in the plot of the specific heat which is offered in Fig. \ref{compspec}.
Nevertheless the plot of $C_v$ vs. $T$ indicates the expected
typical specific heat peak around $T=0.1$ which could not be seen in
the simulations due to the inherent limitations of computer time.
Once the specific heat takes a dive, at some point we can no longer
believe the extrapolation, and more accurate data are necessary to
be able to predict the thermodynamic properties near zero
temperatures.

\section{Summary and the road ahead}
\label{summary}

In summary, we have shown how the scenario of the glass transition
can be neatly characterized by the temperature dependence of the
concentrations of the various quasi-species that are obtained after
up-scaling. Those quasi-species that are readily changed because
they are rich in free energy are of course those that deplete first.
We are left with quasi-species that are low in free energy, and
these would naturally be hard to change since their change necessarily
increases the free energy. The presumed ergodicity allows us to
introduce statistical mechanics which can beautifully encode the
scenario, using a small number of quasi-species together with their
enthalpies and degeneracies. The statistical mechanics applies also
at temperatures that are beyond the range available to molecular
simulations, and can be used to predict what is happening there. We
have shown that when the upscaling and the statistical mechanics
work well, the thermodynamics of the system can be understood very
well using the up-scaled picture.

There are still issues to understand.  One of our main concerns is how to achieve a first-principles prediction of the parameter $\mu$ in Eq. (\ref{fit}). We have, so far, been unable to bridge this gap, leaving the connection between the dynamics and the upscaling theory
still dependent on a phenomenological fit. A satisfactory prediction of $\mu$ would, in our opinion, close the loop and put to rest the riddle of the slowing down in the glass-forming systems. Also, the upscaling approach had been so far limited
to computer model of glass formation, we propose that it would be enlightening to try and apply it to real experimental systems. We hope that these and other issues would be clarified in future research.

\acknowledgments

This work had been supported in part by the Israel Science Foundation, the German Israeli Foundation and the Minerva
Foundation, Munich, Germany.


\begin{thebibliography}{99}
\bibitem{footnote}
Note that in \cite{09LPZ} a smaller set of quasi-species was identified with $C_{\rm liq}$. In the present treatment we
include in this set all the quasi-species whose concentration goes down significantly in the relevant range of tempeartures, and as a result we find a much improved relation between the compositional and dynamical aspects as seen in Fig. \ref{tauvsxi}.
\end{thebibliography}
\end{document}